\documentclass[showpacs,preprintnumbers,amsmath,amssymb]{revtex4}
\usepackage{graphicx}% Include figure files
\usepackage{dcolumn}% Align table columns on decimal point
\usepackage{bm}% bold math
\begin{document}
%\begin{CJK*}{GBK}{song}
\title{Pion Superfluidity beyond Mean Field Approximation In Nambu--Jona-Lasinio Model}
\author{Chengfu Mu, Pengfei Zhuang}
\affiliation{Physics Department, Tsinghua University, Beijing
             100084, China}

\begin{abstract}
We investigate pion superfluidity in the frame of two flavor
Nambu--Jona-Lasinio model beyond mean field approximation. We
calculate the thermodynamics to the next to leading order in an
expansion in the inverse number of colors, including both quark
and meson contributions at finite temperature and baryon and
isospin density. Due to the meson fluctuations, the Sarma phase
which exists at mean field level is washed away, and the
Bose-Einstein condensation region at low isospin density is highly
suppressed.
\end{abstract}

\date{\today}

\maketitle

%%%%%%%%%%%%%%%%%%%%%%%%%%%%%%%%%%%%%%%%%%%%%%%%%%%%%%%%%%%%%%%%%%%%%%%%%
\section {Introduction}
\label{s1}
%%%%%%%%%%%%%%%%%%%%%%%%%%%%%%%%%%%%%%%%%%%%%%%%%%%%%%%%%%%%%%%%%%%%%%%%%
The study on Quantum Chromodynamics (QCD) phase structure is
recently extended to finite isospin density~\cite{son}. The
physical motivation to study QCD at finite isospin density and the
corresponding pion superfluidity is related to the investigation
of compact stars, isospin asymmetric nuclear matter and heavy ion
collisions at intermediate energies.

While the perturbation theory of QCD can well describe the
properties of new QCD phases at extremely high temperature and
density, the study on the phase structure at moderate temperature
and density depends on lattice QCD calculation and effective
models with QCD symmetries. The lattice simulation at finite
isospin chemical potential~\cite{kogut} shows that there is a
phase transition from normal phase to pion superfluidity phase at
a critical isospin chemical potential which is about the pion mass
in the vacuum. The QCD phase structure at finite isospin density
is also investigated in low energy effective models, such as the
Nambu--Jona-Lasinio (NJL) model~\cite{njl} applied to
quarks~\cite{toub,bard,frank,he-1,warringa} which is simple but
enables us to see directly how the dynamic mechanism of isospin
symmetry breaking operates. Near the phase transition point, the
chiral and pion condensates calculated in this model are in good
agreement with the lattice simulation~\cite{kogut}.

In a pion superfluid at zero baryon chemical potential, the quark
and antiquark of a condensed pair have the same isospin chemical
potential and in turn the same Fermi surface. When a nonzero
baryon chemical potential is turned on, it can be regarded as a
Fermi surface mismatch between the quark and antiquark. The pion
superfluidity in baryonic matter is recently discussed at mean
field level in the NJL model in chiral limit~\cite{ebert} and in
real case with finite current quark mass~\cite{he-2}. The pion
superfluid can exist when the baryon density is not very high,
otherwise the system will be in normal phase without pion
condensation because of the too strong mismatch. Inside the pion
superfluid, the condensed state is separated into two phases. At
small isospin chemical potential $\mu_I$, the homogeneous and
isotropic Sarma phase~\cite{sarma} is free from the Sarma
instability~\cite{sarma} and magnetic instability~\cite{wu} due to
the strong coupling and large enough effective quark mass, it is
therefore the stable ground state. At large $\mu_I$, while the
Sarma instability can be cured via fixing baryon density $n_B$ to
be nonzero, its magnetic instability implies that the
inhomogeneous and anisotropic Larkin-Ovchinnikov-Fudde-Ferrell
(LOFF) phase~\cite{loff} is favored than the Sarma phase. In the
intermediate $\mu_I$ region, the stable ground state is the Sarma
phase at higher $n_B$ and LOFF phase at lower $n_B$.

The Bose-Einstein condensation -- Bardeen-Cooper-Shriffer
(BEC-BCS) crossover at finite baryon and isospin chemical
potentials is investigated in the NJL model~\cite{he-3}. The pion
condensation undergoes a BEC-BCS crossover when the isospin
chemical potential increases. The point here is that the crossover
is not triggered by increasing the strength of attractive
interaction among quarks but driven by changing the isospin
density. It is found that the chiral symmetry restoration at
finite temperature and density plays an important role in the
BEC-BCS crossover.

Most of the work in the NJL model is mainly based on the mean
field approximation to the quark mass and on the random phase
approximation (RPA) for the Bethe-Salpeter equation for the meson
masses~\cite{njlquark}. If one examines the thermodynamic
potential in the mean field approximation, one sees immediately
the deficit of this approach, viz, that only the quarks contribute
to the thermodynamic potential with mesons playing no role
whatsoever. This is clearly inadequate and unphysical, since one
expects at least that the pionic degrees of freedom should
dominate the system at low temperature, while the quark degrees of
freedom should be relevant only in the chiral symmetry restoration
phase. As such, this indicates that calculations in the NJL model
must be performed beyond the mean field approximation. In
Ref.~\cite{zhuang}, the thermodynamics of a quark-meson plasma is
calculated to order $1/N_c$ in an expansion in the inverse number
of colors, and pions as Goldstone particles corresponding to
spontaneous chiral symmetry restoration do control the
thermodynamic functions at low temperature and density.

A characteristic feature of the Sarma phase is the intermediate
temperature superfluidity~\cite{liao}: the superfluidity happens
at finite temperature but disappears at zero temperature. Since
the mean field treatment is a good approximation only at zero
temperature~\cite{pieri}, a careful study on the Sarma phase needs
to go beyond the mean field. As for the BEC-BCS crossover induced
by the change in density, the description on the BEC phase at low
density should be closely related to whether the meson
fluctuations are included or not. In this paper, we investigate
the pion superfluidity in the frame of the NJL model beyond mean
field approximation. We will focus on the effect of meson
fluctuations on the Sarma phase and the BEC-BCS crossover at
finite temperature and baryon and isospin density.

The paper is organized as follows. In Section \ref{s2} we present
the thermodynamics of the pion superfluidity and the gap equations
for the chiral and pion condensates in the NJL model in and beyond
mean field approximation. In Section \ref{s3} we calculate the
phase diagram and see the meson effect on the Sarma phase and
BEC-BCS crossover. We summarize and conclude in Section \ref{s4}.

%%%%%%%%%%%%%%%%%%%%%%%%%%%%%%%%%%%%%%%%%%%%%%%%%%%%%%%%%%%%%%%%%%%%%%%%%
\section {Thermodynamics of the Pion Superfluidity}
\label{s2}
%%%%%%%%%%%%%%%%%%%%%%%%%%%%%%%%%%%%%%%%%%%%%%%%%%%%%%%%%%%%%%%%%%%%%%%%%
The two flavor $SU(2)$ NJL Lagrangian density is defined as
\begin{equation}
{\cal L} =
\bar{\psi}\left(i\gamma^{\mu}\partial_{\mu}-m_0+\mu\gamma_0\right)\psi
+G\left[\left(\bar{\psi}\psi\right)^2+\left(\bar{\psi}i\gamma_5{\bf
\tau}\psi\right)^2 \right]
\end{equation}
with scalar and pseudoscalar interactions corresponding to
$\sigma$ and ${\bf \pi}$ excitations, where $\psi$ is the quark
field, $m_0$ the current quark mass, $G$ the coupling constant
with dimension (GeV)$^{-2}$, and $\mu$ the quark chemical
potential matrix in flavor space $\mu=diag(\mu_u,\
\mu_d)=diag(\mu_B/3+\mu_I/2,\ \mu_B/3-\mu_I/2)$ with $\mu_B$ and
$\mu_I$ being baryon and isospin chemical potential. The
Lagrangian density has the symmetry $U_B(1)\bigotimes
SU_I(2)\bigotimes SU_A(2)$ corresponding to baryon number
symmetry, isospin symmetry and chiral symmetry, respectively.
However, at nonzero isospin chemical potential, the isospin
symmetry $SU_I(2)$ breaks down to $U_I(1)$ global symmetry with
the generator $I_3$ which is related to the condensation of
charged pions. At zero baryon chemical potential, the Fermi
surfaces of $u (d)$ and anti-$d (u)$ quarks coincide and hence the
condensate of $u$ and anti-$d$ quarks is favored at sufficiently
high $\mu_I>0$ and the condensate of $d$ and anti-$u$ quarks is
favored at sufficiently high $\mu_I <0$. We introduce the chiral
condensate,
\begin{equation}
\label{chiral}
\sigma=\langle\bar\psi\psi\rangle,
\end{equation}
and the pion condensate,
\begin{equation}
\label{pion}
\pi=\sqrt 2\langle\bar\psi i\gamma_5\tau_+\psi\rangle
= \sqrt 2\langle\bar\psi i\gamma_5\tau_-\psi\rangle
\end{equation}
with $\tau_{\pm} = \left(\tau_1 \pm i\tau_2\right)/\sqrt{2}$. A
nonzero condensate $\sigma$ means spontaneous chiral symmetry
breaking, and a nonzero condensate $\pi$ means spontaneous isospin
symmetry breaking.

In mean field approximation the thermodynamic potential includes
the condensation part and the quark part,
\begin{equation}
\label{omf} \Omega_{mf} = G(\sigma^2+\pi^2)+\Omega_q,
\end{equation}
and the quark part can be evaluated as a summation of four
quasiparticle contributions~\cite{he-2},
\begin{equation}
\label{oq} \Omega_q =-6\sum_{i=1}^4\int {d^3 {\bf p}\over
(2\pi)^3}g(\omega_i),
\end{equation}
where $\omega_i$ are the dispersions of the quasiparticles,
\begin{eqnarray}
\label{omega}
&&\omega_1=E_-+\mu_B/3,\ \ \ \omega_2=E_--\mu_B/3,\nonumber\\
&&\omega_3=E_++\mu_B/3,\ \ \ \omega_4=E_+-\mu_B/3
\end{eqnarray}
with the definitions
\begin{equation}
\label{energy} E_\pm = \sqrt{(E_p\pm\mu_I/2)^2+4G^2\pi^2},\ \ \ \
E_p = \sqrt{p^2+m^2},
\end{equation}
and the function $g(x)$ is defined as $g(x)=x/2+T\ln(1+e^{-x/T})$.
The effective quark mass $m$ is controlled by the chiral
condensate, $m=m_0-2G\sigma$. The gap equations to determine the
condensates $\sigma$ (or quark mass $m$) and $\pi$ can be obtained
by the minimum of the thermodynamic potential
$\Omega_{mf}(T,\mu_B,\mu_I,m,\pi)$,
\begin{equation}
\label{gap1}
\frac{\partial\Omega_{mf}}{\partial m}=0,\ \ \
\frac{\partial\Omega_{mf}}{\partial\pi}=0,\ \ \
{\partial^2\Omega_{mf}\over
\partial m^2}>0,\ \ \ {\partial^2\Omega_{mf}\over\partial \pi^2}>0.
\end{equation}
From the first order derivatives, we have
\begin{eqnarray}
\label{gap2}
&&
m\left({\frac{1}{4G}}+{\frac{\partial\Omega_q}{\partial
m^2}}\right)={\frac{m_0}{4G}},\nonumber\\
&& \pi\left(G+{\frac{\partial\Omega_q}{\partial \pi^2}}\right)=0.
\end{eqnarray}

Considering the relations between $\mu_I,\ \mu_B$ and $\mu_u,\
\mu_{\bar d}$, $\mu_u=\mu_B/3+\mu_I/2$ and $\mu_{\bar
d}=-\mu_B/3+\mu_I/2$, the baryon and isospin density
$n_B=-\partial\Omega_{mf}/\partial\mu_B$ and
$n_I=-\partial\Omega_{mf}/\partial\mu_I$ can be expressed in terms
of the $u$ and $\bar d$ quark density
$n_u=-\partial\Omega_{mf}/\partial\mu_u$ and $n_{\bar
d}=-\partial\Omega_{mf}/\partial\mu_{\bar d}$,
\begin{eqnarray}
\label{number1}
n_I &=& \frac{1}{2}(n_u+n_{\bar d}),\nonumber\\
n_B &=& \frac{1}{3}(n_u-n_{\bar d}).
\end{eqnarray}
It is easy to see that $n_B$ plays the role of density asymmetry
for pion condensation. For isospin symmetric matter with $n_B=0$
the only possible homogeneous and isotropic pion condensed state
is the BCS state. The Sarma state appears only in isospin
asymmetric matter with $n_B\neq 0$. The two gap equations
(\ref{gap2}) and two number equations (\ref{number1}) determine
self-consistently $m, \pi, \mu_I$ and $\mu_B$ as functions of $T,
n_I$ and $n_B$ at mean field level.

We now consider the meson contribution to the thermodynamics of
the system. The meson modes are regarded as quantum fluctuations
above the mean field in the NJL model and can be calculated in the
frame of RPA~\cite{njlquark}. For the mean field quark propagator
with off-diagonal elements in flavor space,
\begin{equation}
{\cal S}^{-1}(p)=\left(\begin{array}{cc} \gamma^\mu p_\mu+\mu_u\gamma_0-m & 2iG\pi\gamma_5\\
2iG\pi\gamma_5 & \gamma^\mu
p_\mu+\mu_d\gamma_0-m\end{array}\right),
\end{equation}
we must consider all possible channels in the bubble summation in
RPA. In the pion superfluidity region, $\sigma$ and charged pions
are coupled to each other and the uncharged pion is decoupled from
them. Using matrix notation for the meson polarization function
$1-2G\Pi(q)$~\cite{he-1},
\begin{equation}
1-2G\Pi=\left(\begin{array}{cccc} 1-2G\Pi_{\sigma\sigma} & -2G\Pi_{\sigma\pi_+} & -2G\Pi_{\sigma\pi_-} & 0\\
                                   -2G\Pi_{\pi_+\sigma} & 1-2G\Pi_{\pi_+\pi_+} & -2G\Pi_{\pi_+\pi_-} & 0\\
                                   -2G\Pi_{\pi_-\sigma} & -2G\Pi_{\pi_-\pi_+} & 1-2G\Pi_{\pi_-\pi_-} & 0\\
                                   0 & 0 & 0 & 1-2G\Pi_{\pi_0\pi_0}\end{array}\right)
\end{equation}
with the quark bubbles
\begin{equation}
\label{bubble} \Pi_{jk}(q) = i\int{d^4 p\over (2\pi)^4} Tr
\left[\Gamma_j^* {\cal S}(p+q)\Gamma_{k} {\cal S}(p)\right],\ \ \
\ j,k=\sigma,\ \pi_+,\ \pi_-,\ \pi_0
\end{equation}
where the trace $Tr = Tr_C Tr_F Tr_D$ is taken in color, flavor
and Dirac spaces and the meson vertexes are defined as
\begin{equation}
\label{vertex} \Gamma_j = \left\{\begin{array}{ll}
1 & j=\sigma\\
i\tau_+\gamma_5 & j=\pi_+ \\
i\tau_-\gamma_5 & j=\pi_- \\
i\tau_3\gamma_5 & j=\pi_0\ ,
\end{array}\right.\ \
\Gamma_j^* = \left\{\begin{array}{ll}
1 & j=\sigma\\
i\tau_-\gamma_5 & j=\pi_+ \\
i\tau_+\gamma_5 & j=\pi_- \\
i\tau_3\gamma_5 & j=\pi_0, \\
\end{array}\right.
\end{equation}
the meson masses $M_j$ are determined by
\begin{equation}
\label{pole} \det\left[1-2G\Pi(q_0+\mu_j=M_j,{\bf q}=0)\right]=0
\end{equation}
with meson chemical potentials $\mu_\sigma=0,\ \mu_{\pi_+}=\mu_I,\
\mu_{\pi_-}=-\mu_I,\ \mu_{\pi_0}=0$.

When the contribution from the meson fluctuations is taken into
account, the total thermodynamic potential to order $1/N_c$ in an
expansion in the inverse number of colors becomes
\begin{equation}
\Omega=\Omega_{mf}+\Omega_{fl},
\end{equation}
where the mean field part $\Omega_{mf}(T,\mu_B,\mu_I,m,\pi)$ is
shown in (\ref{omf}) and the meson part
$\Omega_{fl}(T,\mu_B,\mu_I,m,\pi)$ is expressed in terms of the
polarization function~\cite{zhuang},
\begin{equation}
\label{ofl} \Omega_{fl} = -{\frac{i}{2}}\int{\frac{d^4
q}{(2\pi)^4}}\ln\det\left[1-2G\Pi(q)\right].
\end{equation}
As is expected physically, the mesonic or collective degrees of
freedom play a dominant role at low temperature, while the quark
degrees of freedom are most relevant at high
temperature~\cite{zhuang}.

In the chiral symmetry restoration phase at high temperature
and/or high density, mesons are not stable bound states, but
rather resonant states. They will decay into their quark-antiquark
pairs. As a consequence, the determinant in the logarithm of
(\ref{ofl}) is a complex function in the meson energy plane and
the imaginary part can be expressed as a scattering phase shift
associated with quark-antiquark scattering. From the calculation
in the NJL model with only chiral dynamics~\cite{zhuang}, the
meson width is small around the critical temperature but becomes
remarkable when the meson mass is much larger than two times the
quark mass, and correspondingly, the contribution from the phase
shift to the thermodynamics is negligible at low temperature but
significant when the temperature is high enough. For our
calculation in the pion superfluidity phase, it can be estimated
that the phase shift will be important in the BCS state at high
density but its contribution is weakened in the BEC state at low
density. Since we focus in this paper the Sarma phase and the BEC
state which exist at low isospin density, we take pole
approximation and neglect the scattering phase shift to simplify
the numerical calculations. In pole approximation, the meson
contribution can be greatly simplified as a summation of four
quasiparticles,
\begin{eqnarray}
\label{om} && \Omega_{fl}=\sum_j\Omega_j,\\
&& \Omega_j = \int{\frac{d^3{\bf
q}}{(2\pi)^3}}\left[{\frac{1}{2}}\left(E_j-\mu_j\right)
+T\ln\left(1-e^{-\left(E_j-\mu_j\right)/T}\right)\right]\nonumber
\end{eqnarray}
with meson energies $E_j=\sqrt{M_j^2+{\bf q}^2}$.

While mesons do not change the baryon density of the system, the
charged pions modify the isospin density when the meson
contribution to the thermodynamics is included,
\begin{equation}
\label{number2} n_I={1\over 2}\left(n_u+n_{\bar
d}\right)+\left(n_{\pi_+}-n_{\pi_-}\right),
\end{equation}
where $n_{\pi_+}=-\partial\Omega_{\pi_+}/\partial\mu_{\pi_+}$ and
$n_{\pi_-}=-\partial\Omega_{\pi_-}/\partial\mu_{\pi_-}$ are the
$\pi_+$ and $\pi_-$ density.

Up to this point, the order parameters $m$ for chiral phase
transition and $\pi$ for pion superfluidity have been regarded as
the values minimizing $\Omega_{mf}$, and $\Omega_{fl}$ has been
evaluated at these mean field values, $m=m_{mf}$ and
$\pi=\pi_{mf}$. While this is a correct perturbative expansion
above the mean field, we may ask the questions: What is the
feedback from the mesonic degrees of freedom to the order
parameters and whether we could improve on these mean field values
by regarding $m$ and $\pi$ as variational parameters of the total
thermodynamic potential $\Omega=\Omega_{mf}+\Omega_{fl}$? We now
perform this procedure and see what the difference between the new
and mean field condensates is.

Taking the first order derivatives of the total thermodynamic
potential with respect to the unknown quark mass $m$ and pion
condensate $\pi$, we obtain the following modified gap equations,
\begin{eqnarray}
\label{newgap1} && m\left({1\over
4G}+{\partial\Omega_q\over\partial
m^2}+{\partial\Omega_{fl}\over\partial m^2}\right)={m_0\over
4G},\nonumber\\
&& \pi\left(G+{\frac{\partial\Omega_q}{\partial
\pi^2}}+{\frac{\partial\Omega_{fl}}{\partial\pi^2}}\right)=0.
\end{eqnarray}
In comparison with the mean field gap equations (\ref{gap2}), the
fluctuation part $\Omega_{fl}$ in the thermodynamic potential
leads to a new minimum at $m_{mf+fl}$ and $\pi_{mf+fl}$ that now
differs from the mean field one at $m_{mf}$ and $\pi_{mf}$. It is
easy to see that the structure of the new gap equations guarantees
the two phase transitions. From the second gap equation for pion
superfluidity, the trivial solution $\pi=0$ corresponds to normal
quark matter, while the nonzero solution from the zero of the
bracket corresponds to the energetically favoured pion condensed
state. In the chiral limit, there are also two solutions of the
first gap equation corresponding respectively to the chiral
symmetry breaking and restoration phase.

We now expand the fluctuation part of the thermodynamic potential
around the mean field minimum,
\begin{equation}
\Omega_{fl}(T,\mu_B,\mu_I,m^2,\pi^2) = \sum_{i,j=0}^\infty {1\over
i! j!} {\partial^i\over
\partial (m^2)^i} {\partial^j\over
\partial (\pi^2)^j}\Omega_{fl}(T,\mu_B,\mu_I,m^2,\pi^2)\Big|_{mf}
\left(m^2-m_{mf}^2\right)^i\left(\pi^2-\pi_{mf}^2\right)^j,
\end{equation}
and, to further simplify the calculation, we consider the
expansion only to the first order derivatives. Inserting the
expansion into the new gap equations yields the following gap
equations
\begin{eqnarray}
\label{newgap2}
&& m\left({1\over
4G_\sigma}+{\partial\Omega_q\over\partial m^2}\right)={m_0\over
4G},\nonumber\\
&& \pi\left(G_\pi+{\frac{\partial\Omega_q}{\partial
\pi^2}}\right)=0
\end{eqnarray}
with two effective coupling constants $G_\sigma$ and $G_\pi$
defined by
\begin{eqnarray}
\label{gspi} && {1\over 4G_\sigma}={1\over 4G}+{\partial\over
\partial
m_{mf}^2}\Omega_{fl}(T,\mu_B,\mu_I,m_{mf}^2,\pi_{mf}^2),\nonumber\\
&& G_\pi = G+{\partial\over
\partial
\pi_{mf}^2}\Omega_{fl}(T,\mu_B,\mu_I,m_{mf}^2,\pi_{mf}^2).
\end{eqnarray}
In comparing this group of coupled gap equations with the mean
field one (\ref{gap2}), one observes that, in the chiral limit,
the two groups take the same form, differing only in the effective
coupling constants. The coupling constants in the scalar and
pseudoscalar channels are the same at mean field level, but they
become different and depend on temperature and charge densities
when one goes beyond the mean field. If we take
$G_\sigma=G_\pi=G$, we recover the mean field case. That is, in
this approach, the contribution from meson fluctuations is fully
included in $G_\sigma$ and $G_\pi$.

The above approach describes the thermodynamics of a quark-meson
plasma with both chiral phase transition and pion superfluidity
phase transition beyond the mean field at finite temperature and
baryon and isospin density. The two new gap equations
(\ref{newgap2}) determine simultaneously the order parameters
$\sigma$ (or $m$) and $\pi$ of the two phase transitions. In the
chiral limit, the two phase transitions are fully separated from
each other~\cite{he-1}: the chiral symmetry is automatically
restored in the pion superfluidity phase. That is, the two order
parameters do not coexist in the system. In the real word, chiral
symmetry is not fully restored at any isospin chemical potential.
However, $\sigma$ is much smaller than $\pi$ in the pion
superfluidity region~\cite{he-1}. Since we focus in this paper on
the fluctuation effect on the pion superfluidity, we will, for the
purpose of simplification in numerical calculations, neglect the
$\sigma$ fluctuations and keep only the $\pi$ fluctuations in the
gap equations. Namely, we take $G_\sigma=G$ in the following.

%%%%%%%%%%%%%%%%%%%%%%%%%%%%%%%%%%%%%%%%%%%%%%%%%%%%%%%%%%%%%%%%%%%%%%%%%
\section {Phase diagrams in and beyond mean field}
\label{s3}
%%%%%%%%%%%%%%%%%%%%%%%%%%%%%%%%%%%%%%%%%%%%%%%%%%%%%%%%%%%%%%%%%%%%%%%%%
Since the NJL model is non-renormalizable, we should employ a
regularization scheme to avoid the divergence in the gap
equations. The simplest and normally used way is to introduce a
hard three momentum cutoff $|{\bf p}|<\Lambda$. In the following
numerical calculations, we take the current quark mass $m_0=5$
MeV, the coupling constant $G=4.93$ GeV$^{-2}$ and the cutoff
$\Lambda=653$ MeV~\cite{zhuang}. This group of parameters ensures
the pion mass $m_\pi=138$ MeV and the pion decay constant
$f_\pi=93$ MeV in the vacuum.

\begin{figure}[htbp]
\includegraphics[width=8cm]{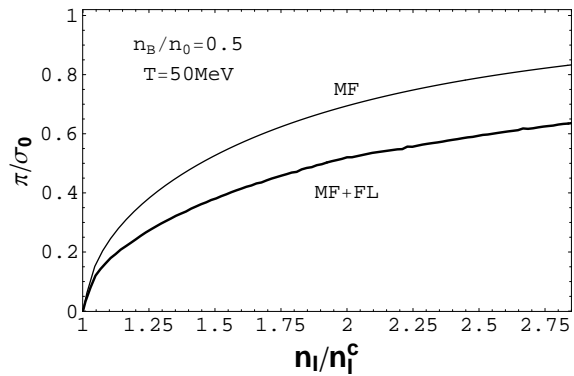}
\caption { The pion condensates in mean field approximation (thin
line) and including meson fluctuations (thick line) as functions
of isospin density at fixed temperature and baryon density.
$\sigma_0$ is the chiral condensate in vacuum and $n_I^c$ is the
critical isospin density of pion superfluidity. } \label{fig0}
\end{figure}

In the treatment above, we considered the meson fluctuations as a
perturbation around the mean field and took only the first order
derivatives in the effective coupling constants (\ref{gspi}). If
this treatment is good, the difference between the two pion
condensates calculated in and beyond the mean field approximation
should be small. To check the validity region of this method, we
show in Fig.\ref{fig0} the two condensates as a function of
isospin density at fixed temperature and baryon density. At low
isospin density which corresponds to the BEC region, the
difference between the two is really small, but it grows with
increasing density and becomes large in the BCS region. Therefore,
the approximation with only first order derivatives is good for
the study of BEC, but the contribution from the higher order
derivatives may be important for the BCS state.
%%%%%%%%%%%%%%%%%%%%%%%%%%%%%%%%%%%%%%%%%%%%%%%%%%%%%%%%%%%%%%%%%%%%%%%
%%%%%%%%%%%%%%%%%%%%%%%%%%%%%%%%%%%%%%%%%%%%%%%%%%%%%%%%%%%%%%%%%%%%%%%

\begin{figure}[htbp]
\includegraphics[width=8cm]{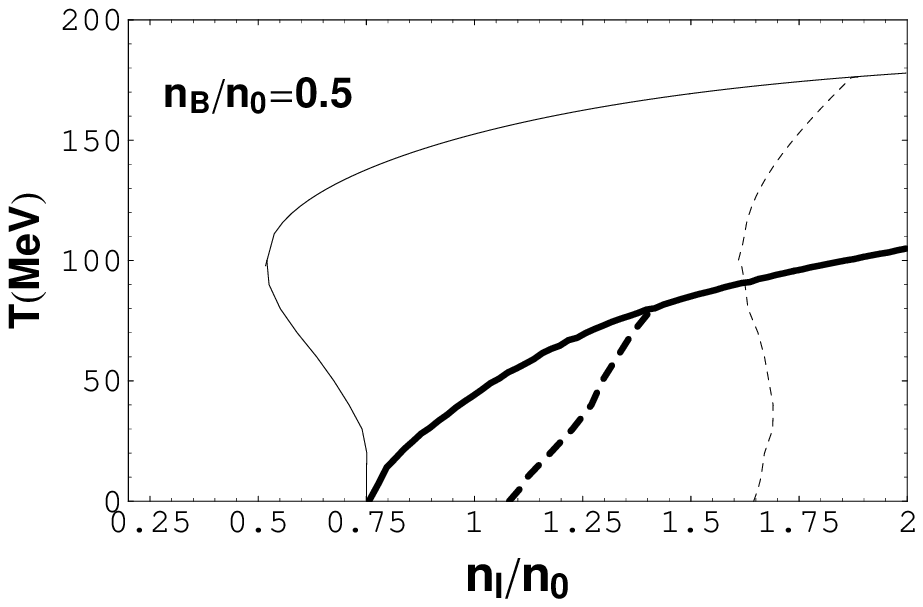}
\includegraphics[width=8cm]{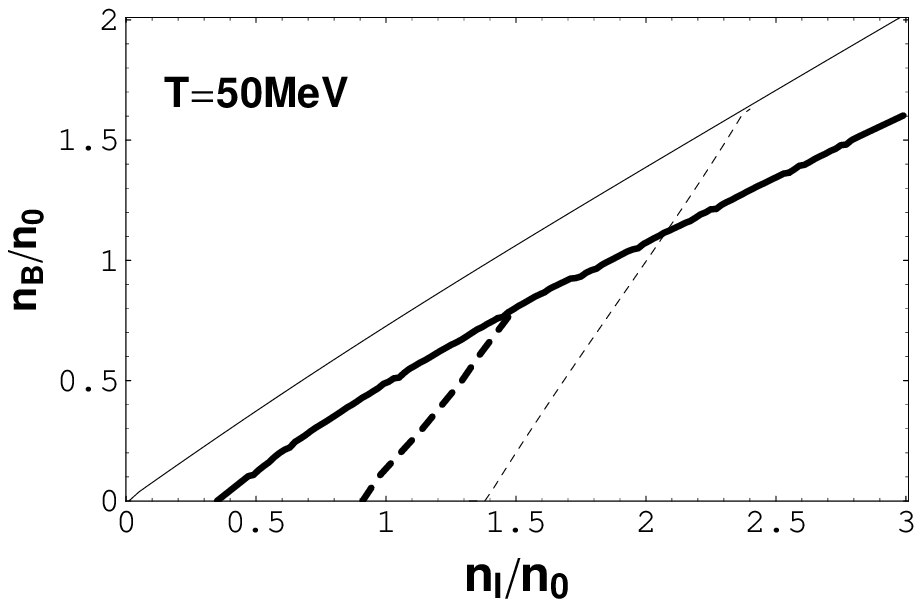}
\caption { The phase diagrams of pion superfluidity in $T-n_I$
plane at fixed baryon density $n_B/n_0=0.5$(left panel) and in
$n_B-n_I$ plane at fixed temperature $T=50$ MeV(right panel) in
mean field approximation (thin lines) and including meson
fluctuations (thick lines). $n_0=0.17/fm^3$ is the normal nuclear
density. The solid lines are the phase transition lines and the
dashed lines are the BEC-BCS crossover lines. } \label{fig1}
\end{figure}

The phase diagrams of pion superfluidity in $T-n_I$ plane at fixed
baryon density and in $n_B-n_I$ plane at fixed temperature are shown
in Fig.\ref{fig1}. The thin and thick solid lines are respectively
phase transition lines in and beyond mean field approximation which
separate the normal quark matter at high temperature or high baryon
density from the pion superfluidity matter at high isospin density.
For pion superfluidity, the averaged Fermi surface of the paired
quarks is controlled by isospin chemical potential and the mismatch
is served by baryon chemical potential. Therefore, the Sarma phase
which is induced by the Fermi surface mismatch may enter the pion
superfluidity at nonzero baryon density. In mean field
approximation, by analyzing the four quasiparticle dispersions
$\omega_i$, the possible types of Sarma state and their
thermodynamic and dynamic instabilities are discussed in detail in
Ref.\cite{he-1}. It is found that the Sarma phase is the ground
state of the pion superfluidity at low isospin chemical potential.
Very different from the BCS phase structure where the temperature of
the pairing state is always lower than the temperature of the normal
state, the Sarma phase appears in an intermediate temperature region
and the normal state exists in lower and higher temperature regions,
see the mean field phase transition line in the $T-n_I$ plane at low
isospin density in Fig.\ref{fig1}. However, the phase structure in
the $T-n_I$ plane is significantly modified when the meson
fluctuations are included. From Fig.\ref{fig1}, the meson effect
reduces greatly the pion superfluidity region, and the critical
temperature is suppressed from about 150 MeV in mean field treatment
to about 80 MeV in the case beyond the mean field. A qualitative
change resulted from the meson fluctuations is that the intermediate
temperature superfluidity or the Sarma state in mean field
calculations is totally washed away, and the normal quark matter is
always above the BCS pairing state. In the $n_B-n_I$ plane, the
phase diagram in mean field approximation is similar to the one in
$n_B-\mu_I$ plane obtained in Ref.\cite{he-2}, and again the meson
effect reduces remarkably the pion superfluidity region.
%%%%%%%%%%%%%%%%%%%%%%%%%%%%%%%%%%%%%%%%%%%%%%%%%%%%%%%%%%%%%%%%%%%%%%%
%%%%%%%%%%%%%%%%%%%%%%%%%%%%%%%%%%%%%%%%%%%%%%%%%%%%%%%%%%%%%%%%%%%%%%%

\begin{figure}[htbp]
\includegraphics[width=8cm]{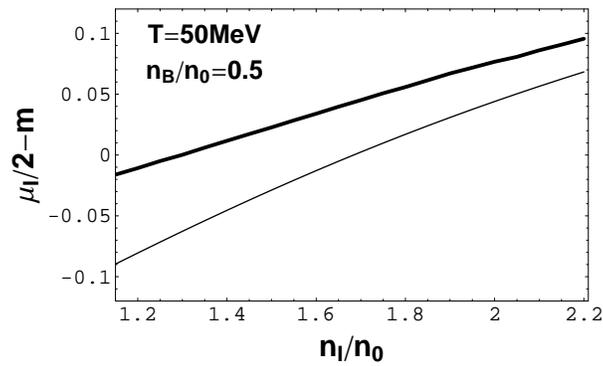}
\caption { The effective chemical potential $\mu_I/2-m$ as a
function of isospin density at fixed temperature $T=50$ MeV and
baryon density $n_B/n_0=0.5$ in mean field approximation (thin
line) and including mesonic fluctuations (thick line). $n_0$ is
the normal nuclear density.} \label{fig2}
\end{figure}

It has been argued both in effective theory and lattice simulation
that at finite but not very large isospin density and zero baryon
density, the QCD matter is a pure meson matter, i.e., a
Bose-Einstein condensate of charged pions. At ultrahigh isospin
density, the matter turns to be a Fermi liquid with
quark-antiquark cooper pairing\cite{son}. Therefore, there should
be a BEC to BCS crossover when the isospin chemical potential
increases. There are some equivalent quantities to describe the
BEC-BCS crossover induced by changing charge density~\cite{mao}.
Among them are the root-mean-square radius of the Cooper pair
which is small in BEC and large in BCS, the s-wave scattering
length which is positive in BEC and negative in BCS, the
condensate scaled by the Fermi energy which is large in BEC and
small in BCS, and the fermion chemical potential which is negative
in BEC and positive in BCS. In the following we take the chemical
potential to characterize the BEC-BCS crossover. For relativistic
pion superfluidity, the chemical potential which controls the
BEC-BCS crossover is $\mu_I/2-m$~\cite{he-3} depending on
temperature and baryon density through the effective quark mass
$m$, and $\mu_I$ can be viewed as the binding energy of the bound
state of quark and antiquark in the BEC limit. In Fig.\ref{fig2}
we show $\mu_I/2-m$ as a function of $n_I$ at fixed temperature
and baryon density in and beyond mean field approximation. In both
cases, the effective chemical potential goes up from negative to
positive values with increasing isospin density. The zero point,
namely the BEC-BCS crossover point, is located at $n_I/n_0=1.68$
in mean field treatment and $n_I/n_0=1.29$ in the case with meson
fluctuations. The crossover lines determined by $\mu_I/2-m=0$ in
$T-n_I$ and $n_B-n_I$ planes are shown in Fig.\ref{fig1}. When the
mesonic fluctuations are included, not only the pion superfluidity
region is greatly reduced, but also the BEC region is strongly
shrunk.
%%%%%%%%%%%%%%%%%%%%%%%%%%%%%%%%%%%%%%%%%%%%%%%%%%%%%%%%%%%%%%%%%%%%%%%
%%%%%%%%%%%%%%%%%%%%%%%%%%%%%%%%%%%%%%%%%%%%%%%%%%%%%%%%%%%%%%%%%%%%%%%

\begin{figure}[htbp]
\includegraphics[width=8cm]{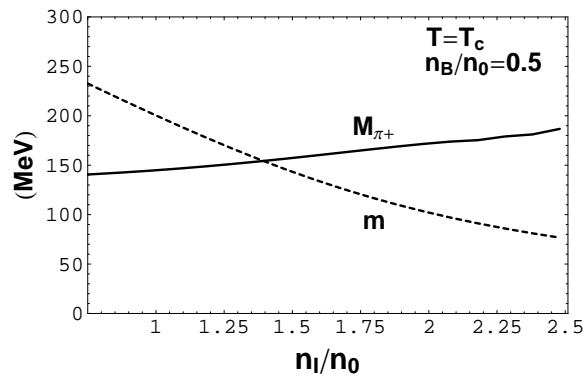}
\caption {The meson mass $M_{\pi_+}$ (solid line) and quark mass
$m$ (dashed line) as functions of isospin density at the critical
temperature $T_c$ and fixed baryon density $n_B/n_0=0.5$. $n_0$ is
the normal nuclear density.} \label{fig3}
\end{figure}

In the BCS limit of the pion superfluidity, the isospin density is
high and the paired quark and antiquark is weakly coupled. At the
critical temperature, the condensate disappears and the weakly
coupled fermions are excited separately, and the system is a Fermi
liquid. In the BEC limit, however, the isospin density is low and
the paired quark and antiquark is tightly coupled. In this case,
above the critical temperature, the system becomes a Bose liquid
of tightly bound pions, and the quarks should be too heavy to be
excited. This means that, at the critical temperature mesons are
lighter than quarks in the BEC limit and quarks are lighter than
mesons in the BCS limit. To confirm the BEC-BCS crossover picture
obtained above by calculating the effective chemical potential
inside the pion superfluidity, we show in Fig.\ref{fig3} the meson
mass $M_{\pi_+}$ and quark mass $m$ as functions of isospin
density at the critical temperature and fixed baryon density. With
increasing isospin density, the quark mass drops down but the
meson mass goes up monotonously. The two lines cross at about
$n_I/n_0=1.4$ which is qualitatively in agreement with the BEC-BCS
crossover value determined by $\mu_I/2-m=0$.
%%%%%%%%%%%%%%%%%%%%%%%%%%%%%%%%%%%%%%%%%%%%%%%%%%%%%%%%%%%%%%%%%%%%%%%
%%%%%%%%%%%%%%%%%%%%%%%%%%%%%%%%%%%%%%%%%%%%%%%%%%%%%%%%%%%%%%%%%%%%%%%%%

%%%%%%%%%%%%%%%%%%%%%%%%%%%%%%%%%%%%%%%%%%%%%%%%%%%%%%%%%%%%%%%%%%%%%%%%%
\section{Summary}
\label{s4}
%%%%%%%%%%%%%%%%%%%%%%%%%%%%%%%%%%%%%%%%%%%%%%%%%%%%%%%%%%%%%%%%%%%%%%%%%
We have investigated the thermodynamics of a pion superfluid at
finite isospin density in the frame of two flavor NJL model beyond
the mean field approximation. Considering the fact that mesons, in
particular pions because of their low mass, dominate the
thermodynamics of a quark-hadron system at low temperature, the
mesonic fluctuations should be significant for the phase structure
of pion superfluidity. By recalculating the minimum of the
thermodynamic potential including meson contribution, we derived a
new gap equation for the pion condensate which is similar to the
mean field form but with a medium dependent coupling constant.
From our numerical calculations, the main effects of the meson
fluctuations on the phase structure are: 1) the critical
temperature of pion superfluidity is highly suppressed and the
Sarma phase which exists at low isospin chemical potential in mean
field approximation is fully washed away, and 2) the BEC region at
low isospin density is significantly shrunk.

{\bf Acknowledgement:} The work is supported by the NSFC Grant
10735040 and the National Research Program Grants 2006CB921404 and
2007CB815000.

%\newpage
%\end{CJK*}
\end{document}